\RequirePackage{fix-cm}
\documentclass[smallextended,final,natbib]{svjour3}       
\smartqed  

\usepackage[T1]{fontenc}
\usepackage[utf8]{inputenc}

\usepackage[final]{graphicx}
\usepackage{bm}
\usepackage{listings}
\usepackage{subfigure}
\usepackage{amsmath}
\usepackage{amssymb}
\usepackage{booktabs}
\usepackage{mathptmx}
\usepackage{enumitem}
\usepackage{hyperref}
\usepackage{epstopdf}
\hypersetup{colorlinks,citecolor=blue}

\begin{document}
	
	\title{Positive Stochastic Collocation for the Collocated Local Volatility Model
	}
	
	
	\author{Fabien {Le Floc'h} \and Cornelis W. Oosterlee 
	}
	
	
	\institute{Fabien Le Floc'h \at
		Delft Institute of Applied Mathematics, TU Delft, Delft, The Netherlands \\
		\email{f.l.y.lefloch@tudelft.nl}           
		\and Cornelis W. Oosterlee, CWI-Centrum Wiskunde \& Informatica, Amsterdam, The Netherlands
	}
	
	\date{Received: date / Accepted: date}

	\maketitle
	
	\begin{abstract}
		This paper presents how to apply the stochastic collocation technique to assets that can not move below a boundary. It shows that the polynomial collocation towards a lognormal distribution does not work well. Then, the potentials issues of the related collocated local volatility model (CLV) are explored. Finally, a simple analytical expression for the Dupire local volatility derived from the option prices modelled by stochastic collocation is given.
	\end{abstract}

\newcommand{\sgn}{\mathop{\mathrm{sgn}}}

\section{Introduction}
In the standard Black-Scholes model, the underlying asset follows a geometrical Brownian motion \citep{black1973pricing}. The model is typically applied to obtain the prices and hedges of financial derivative contracts, such as options on a foreign exchange rate, a stock price or a swap rate. For a given maturity date, the market price of a vanilla option for each quoted strike does not match the assumption of a constant Black-Scholes volatility. The volatilities implied by the market exhibit a smile. This is also true in the maturity dimension, where the market implies a particular term-structure of volatilities. Furthermore, the implied volatilities vary in time. 

In order to solve this discrepancy, \citet{dupire1994pricing} proposed a model where the volatility is made local: it is a function of the asset price and a particular time. This model however requires a smooth and continuous arbitrage-free Black-Scholes implied volatility representation accross the time and asset price dimensions. Finding a good representation is the principal challenge of this model. The local volatility model also suffers from an unrealistic dynamic of the smile in time \citep{hagan2002managing}. In practice this means that forward starting options are mispriced under the local volatility model.

A different approach is to assume that the volatility is stochastic. Several stochastic volatility models that retain some analytical and numerical tractability have been explored over the years, the most popular being the model from \citet{heston1993closed}. They all suffer from similar issues: they don't allow to match the implied volatility smile for short maturities very well and they can be challenging to calibrate properly. A fix for the former issue is to mix stochastic and local volatilities together \citep{alexander2004hedging} at the cost of increasing the computational complexity, which involves either an iterative partial differential equation technique as in \citep{ren2007calibrating} or the particle method for Monte-Carlo \citep{guyon2011smile, van2014heston}. A good local volatility representation is a prerequisite for the stochastic local volatility model.

\citet{grzelak2016clv} proposes another alternative with the collocated volatility model (CLV), where the model prices are calibrated to the market options with the stochastic collocation technique, used as a convenient representation of the terminal distribution. A specific dynamic is added in the form of a stochastic driver process, which allows more control over the prices of forward starting options. This is reminiscent of the Markov functional model of \cite{hunt2004financial}, initially derived for interest rate models and extended to equity models in \citep{fries2006markov}. How to implement this model for equity derivatives in practice? What are its limitations? Those are a few key questions that we attempt to answer in this paper.

The outline of the paper is as follows. Section \ref{sec:overview_collocation_clv} introduces the stochastic collocation technique and the CLV model with a Gaussian driver process with time-dependent volatility. We pay attention to explain its relationship with the Markov functional model. Section \ref{sec:positive_collocation} extends the collocation technique  described in \citep{lefloch2019model1} to assets that can not move below a boundary. For example, a stock price must stay positive. This is particularly important as many equity derivative contracts are written on a future performance, and we show that the price of such contracts can not be estimated by a model that allows negative asset prices. Furthermore, we find that the polynomial collocation towards a lognormal distribution does not work well in practice.
Section \ref{sec:calibration_autocorrelation_clv} presents how to calibrate the autocorrelation such that the CLV model stays compatible with the assumption of deterministic interest rates, in a similar fashion as \citet{jackel2005practical, brockhaus2006implied} do with the Markov functional model. Finally, in section \ref{sec:local_volatility_collocation}, we show that, with a specific choice of interpolation in time, the stochastic collocation leads to a simple analytical expression for the Dupire local volatility.

\section{Stochastic collocation and the collocated volatility model}\label{sec:overview_collocation_clv}
\subsection{Overview of the stochastic collocation method}\label{sec:overview_collocation}
The stochastic collocation method \citep{mathelin2003stochastic} consists in mapping a physical random variable $Y$ to a point $X$ of an artificial stochastic space. Collocation points $x_i$ are used to approximate the function mapping $X$ to $Y$, $F_X^{-1} \circ F_Y$, typically by a polynomial, where $F_X, F_Y$ are respectively the cumulative distribution functions (CDF) of $X$ and $Y$. Thus only a small number of inversions of $Y$ (and evaluations of $F_Y$) are used. This allows the problem to be solved in the "cheaper" artificial space.

In the context of option price interpolation, the stochastic collocation will allow us to interpolate the market CDF in a better set of coordinates. In particular, we will follow \citet{grzelak2017arbitrage} and use a Gaussian distribution for $X$.

In \citep{grzelak2017arbitrage}, the stochastic collocation is applied to the survival distribution function $G_Y$, where $G_Y(y) = 1 - F_Y(y)$ with $F_Y$ being the cumulative density function of the asset price process. 
When the survival density function is known for a range of strikes, their method can be summarized by the following steps:
\begin{enumerate}
	\item Given a set of collocation strikes $y_i$, $i=0,...,N$, compute the survival density $p_i$ at those points: $p_i = G_Y(y_i)$.
	\item Project on the Gaussian distribution by transforming the $p_i$ using the inverse cumulative normal distribution $\Phi^{-1}$ resulting in $x_i = \Phi^{-1}(1-p_i)$.
	\item Interpolate $(x_i,y_i)$ with a Lagrange polynomial $g_N$.
	\item Price by integrating the density with the integration variable $x=\Phi^{-1}(1-G_Y(y))$, using the Lagrange polynomial for the transform.
\end{enumerate}
Let us now detail the last step. The undiscounted price of an option of strike $K$ is obtained by integrating over the probability density, with a change of variables,
\begin{align}
C(K) &= \int_{0}^{+\infty} \max\left(y-K, 0\right) f(y) dy \\ 
&= \int_{\Phi^{-1}(0)}^{\Phi^{-1}(1)} \max\left(G_Y^{-1}(1-\Phi(x))-K, 0\right) \phi(x) dx\nonumber\\
&\approx \int_{-\infty}^{\infty} \max\left(g_N(x)-K,0\right) \phi(x) dx\nonumber\\
\label{eqn:collocation_call_integral}     &= \int_{c_K}^{\infty} (g_N(x)-K) \phi(x) dx\,, 
\end{align}
where $\phi(x)$ is the Gaussian density function and $f$ the probability density implied by the options prices and
\begin{equation}
c_K = g_N^{-1}(K)\,. \label{eqn:collocation_ck}
\end{equation}
The change of variables is valid when the survival density is continuous and its derivative is integrable. In particular, it is not necessary for the derivative to be continuous.
 
As shown in \citep{hunt2004financial}, a polynomial multiplied by a Gaussian can be integrated analytically as integration by parts leads to a recurrence relationship on $m_i(b) = \int_{b}^{\infty} x^i \phi(x) dx$. This idea is also the basis of the Sali tree method \citep{hu2006cutting}. The recurrence is given by
\begin{align}\label{eqn:recurrence}
  m_{i+2}(b) = (i+1) m_i(b) + b^{i+1} \phi(b)   \,,
\end{align}
with $m_0(b) = \Phi(-b), m_1(b) = \phi(b)$.
We have then:
\begin{equation}
C(K) = \sum_{i=0}^{N} a_i m_i(c_K)  - \Phi(-c_K) K\,,
\end{equation}
where $a_i$ are the coefficients of the polynomial in increasing powers.

The terms $m_i(K)$ involve only $\phi(c_K)$ and $\Phi(-c_K)$. The computational cost for pricing one vanilla option can be approximated by the cost of finding $c_K$ and the cost of one normal density function evaluation plus one cumulative normal density function evaluation. For cubic polynomials, $c_K$ can be found analytically through Cardano's formula \citep{nonweiler1968algorithms} and the cost is similar to the one of the Black-Scholes formula. In the general case of a polynomial $g_N$ of degree $N$, the roots can be computed in $O(N^3)$ as the eigenvalues of the associated Frobenius companion matrix $M$ defined by
\begin{equation*}
M(g_N) = 
\begin{pmatrix}
0 & 0 & \cdots & 0 & -\frac{a_0}{a_N}\\
1 & 0 &\cdots & 0 & -\frac{a_1}{a_N}\\
0 & 1  & & 0 & -\frac{a_2}{a_N}\\
\vdots &  \vdots & \ddots & \vdots & \vdots\\
0 & 0 & \cdots  & 1 & -\frac{a_{N-1}}{a_N}\\
\end{pmatrix}\,.
\end{equation*}
We have indeed $\det \left(\lambda I -M\right) = g_N(\lambda)$.
This is, for example, how the Octave or Matlab \texttt{roots} function works \citep{moler1991cleve}. Note that for a high degree $N$, the system can be very ill-conditioned. A remedy is to use a more robust polynomial basis such as the Chebyshev polynomials and compute the eigenvalues of the colleague matrix \citep{good1961colleague,trefethen2011six}. Jenkins and Traub solve directly the problem of finding the roots of a real polynomial in \citep{jenkins1975algorithm}.

A simple alternative, particularly relevant in our case as the polynomial needs to be invertible and thus monotonic, is to use the third-order Halley's method \citep{gander1985halley} with a simple initial guess $c_K=-1$, if $K<F(0,T)$, or $c_K=1$, if $K \geq F(0,T)$, with $F(0,T)$ the forward price to maturity $T$. In practice not more than three iterations are necessary to achieve an accuracy around machine epsilon.

The put option price is calculated through the put-call parity relationship, namely
\begin{equation*}
C(K)-P(K) = F(0,T)-K\,,
\end{equation*}
where $P(K)$ is the undiscounted price today of a put option of maturity $T$, and $F(0,T)$ is the forward price to maturity.

\subsection{Collocated Local Volatility with a Gaussian driver process}
We consider a dynamic model where 
\begin{align}
X(t) &= \sigma_X(t) W(t)\,,\\
S(t) &= g(t,X(t))\,,
\end{align}
with $W$ a Brownian motion and $g$ a collocation function.

Now, from \cite{lefloch2019model1}, we know how to calibrate the collocation polynomial $g_N(t_i, z)$ for each maturity independently, on a Gaussian random variable $Z$ with mean zero and unit standard deviation. The polynomials can be reused in the Monte-Carlo simulation.

Each path followed by the asset $S$ at times $(t_i)_{0<i\leq m}$ in the  Monte-Carlo simulation can be computed without any discretization bias, as follows:
\begin{itemize}
	\item Draw $m$ independent uniform random numbers $U_i$.
\item	Transform them to Gaussian random numbers $Z_i = \Phi^{-1}(U_i)$.
\item	Compute $X_i = X_{i-1} +  Z_i \sqrt{\int_{t_{i-1}}^{t_i} \sigma_X(v)^2 dv}$, starting at $X_0=0, t_0=0$.
\item Compute $S(t_i) = g_N(t_i, \frac{X(t_i)}{\sqrt{\int_{0}^{t_i} \sigma_X(v)^2 dv}})$.
\end{itemize}

The underlying idea is not new. It has been proposed by \citet{jackel2005practical} and \citet{brockhaus2006implied}. Instead of using the collocation on a Gaussian random variable technique, they derive the functional $g$ from the risk-neutral density, and do not specify a way to imply this risk neutral density. This is often referred to as the Markov functional model, initially derived for interest rate models \citep{hunt2004financial}
and extended to equity models in \citep{fries2006markov}. Note that the equity Markov functional model uses the equity as the numeraire, so it is not strictly the same. 
In the CLV model, the functional is directly calibrated to the market.

\section{Positive collocation}\label{sec:positive_collocation}
\begin{figure}[h]
	\centering{
		\includegraphics[width=0.95\textwidth]{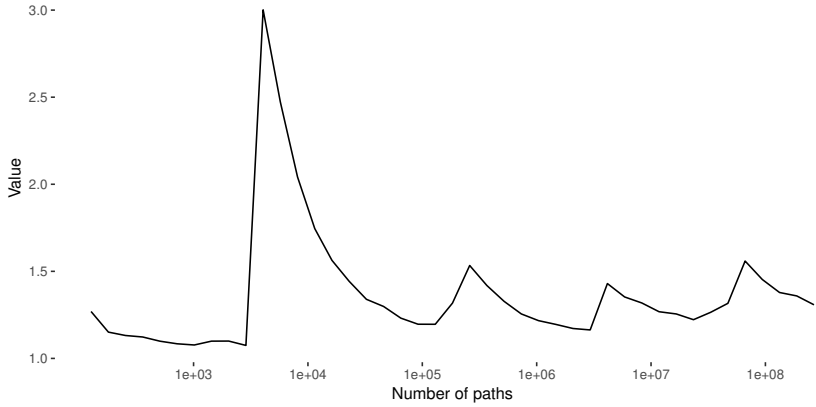}
		\caption{Value of a contract on the future performance obtained by Monte-Carlo simulation of a calibrated CLV model, as a function of the number of paths.\label{fig:tsla_poly_ratio_convergence}}}
\end{figure}

In many applications, the asset can not become negative. But when we apply the stochastic collocation towards a Gaussian variable $X$, there is a non-zero probability that the asset $S$ becomes negative.
For example, many exotic equity derivative contracts are written on the future performance $\frac{S(t_i)}{S({t_j})}$ of the asset. This performance can be computed in a meaningful way only if the asset can not become negative. For example, let us consider the simplest case, a contract paying a future performance $\frac{S(t_2)}{S({t_1})}$ at $t_2$, with $t_1 < t_2$. Its value $V$ at time $t=0$ is \begin{equation}
V(0)=\mathbb{E}_{\mathbb{Q}}\left[\frac{S(t_2)}{S({t_1})}\right]\,.
\end{equation}
Even if the probability that $S(t_1)$ becomes negative or zero is extremely small, this expectation is not well defined. In practice, when the CLV model with a Gaussian driver of constant volatility is calibrated to three distinct maturities of options on the TSLA stock (table \ref{tbl:tsla_poly_coeff}), the mean value of a Monte-Carlo simulation of this contract will jump randomly with the number of paths (Figure \ref{fig:tsla_poly_ratio_convergence}). The standard error never converges to zero.
\begin{table}[h]
\centering{
		\caption{Coefficients of collocation polynomials calibrated to TSLA options on June 15, 2018.\label{tbl:tsla_poly_coeff}}
	\begin{tabular}{rrrrrrr}\toprule
		Expiry & $a_0$& $a_1$& $a_2$&   $a_3$ & $a_4$ & $a_5$ \\ \midrule
	July 20, 2018 & 356.64 & 48.632 & 0.842 & -0.565 & 0.0917 & 0.412\\
	January 18, 2019 &362.86 & 117.77& -23.49 & 3.970 &  5.586 & 0.729\\
	January 17, 2020 &364.01 & 216.74& -72.76& -29.51 & 21.83 & 7.014\\ \bottomrule
	\end{tabular}}
\end{table}


\subsection{Absorption}
 One remedy is to consider that the density accumulates at $S=L$ into a discrete probability mass $p_a$ as in \citep{grzelak2018stochastic}. We obtain $p_a = \Phi(c_L)$, with $c_L = g_N^{-1}(L)$ and the total density over $[L, +\infty)$ will then still sum to one. The asset will not take any value below $L$. In order to ensure strict positivity, we can consider a small but positive $L$. 
It is possible impose the monotonicity of the collocation polynomial and to conserve the first moment in the calibration, while at the same time taking into account this probability mass, by following the technique described
in \citep{lefloch2019model1}. The first moment is
\begin{align}
F(0,T) &= L p_a + \int_{c_L}^{\infty} g_N(x) \phi(x)dx\nonumber\\
&=L \Phi(c_L)+\sum_{i=0}^{N} a_i m_i(c_L)\,,
\end{align}
with $m_i$ defined in equation (\ref{eqn:recurrence}). In order to imply $a_0$ from $(a_i)_{i \geq 1}$ and $F(0,T)$, a one-dimensional non-linear solver is required as $c_L$ is a function of the coefficient $a_0$. The $a_0$ obtained by integrating from $-\infty$ to $\infty$ provides a good initial guess.

In order to price European options, we simply integrate the payoff on $[L, \infty)$ instead of $(-\infty,\infty)$. The call option value is the same as with the standard collocation technique on the polynomial $g_N$, for strikes $K \geq L$. For $K < L$, the undiscounted call option value is the intrinsic value $F(0,T)-K$.
\begin{figure}[h]
	\centering{
		\includegraphics[width=0.95\textwidth]{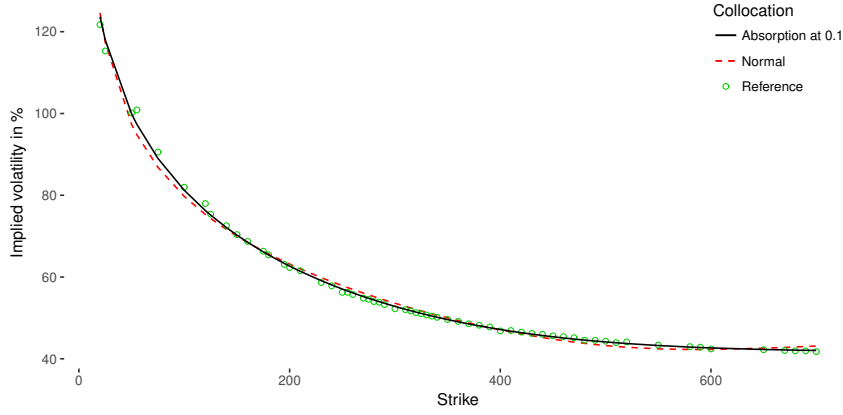}
		\caption{Implied volatility obtained from the calibrated quintic collocation polynomial with absorption for TSLA options expiring on January 17, 2020 as of June 15, 2018.\label{fig:tsla_180615_200117_polyabs}}}
\end{figure}
The resulting fit in terms of implied volatilities is very good, very similar to the standard collocation without absorption. Figure \ref{fig:tsla_180615_200117_polyabs} shows the implied volatility when the collocation polynomial is calibrated to TSLA options expiring on January 17, 2020 as of June 15, 2018. 

In the collocated local volatility model (CLV) of \citet{grzelak2016clv}, a dynamic is added to $X$, for example $X$ is a Wiener process and not just a Gaussian random variable. In a Monte-Carlo simulation of the CLV model with absorption, each time-step is effectively computed independently of each other and the probability of absorption is already taken into account implicitly in the function $g_N$. Unlike a more standard Monte-Carlo simulation of a process with absorption, there is no need to take into account of the eventual absorption at prior times $t_i< t_j$ for a given observation date $t_j$.

Another technique to include absorption is to apply the reflection method \citep[p. 220]{john1982partial} to the probability density implied by the stochastic polynomial collocation. In this case, the probability of absorption becomes 
\begin{equation}
p_a = 2 \Phi(c_L)
\end{equation}
and the first moment is
\begin{align}
F(0,T) &= L p_a + \int_{c_L}^{\infty} g_N(x)\phi(x)dx -\int_{-\infty}^{c_L} (2L-g_N(x))\phi(x)dx\nonumber\\
&= \int_{-\infty}^{\infty} g_N(x)dx\nonumber\\
&=  a_0 + \sum_{i=1}^{\frac{N-1}{2}}   a_{2i} (2i-1)!!\,.
\end{align}
In particular, it corresponds to the first moment of the standard polynomial collocation on $\mathbb{R}$.
Under this absorption model, the undiscounted Call option price is
\begin{equation}
C(K) \approx \int_{c_L}^{\infty} \max\left(g_N(x)-K,0\right)\phi(x) dx - \int_{-\infty}^{c_L} \max\left(2L-g_N(x)-K, 0\right) \phi(x) dx\,.
\end{equation}
Assuming that $K > L$, we have
\begin{align}
C(K) &\approx \int_{c_K}^{\infty} \left(g_N(x)-K\right) \phi(x) dx - \int_{-\infty}^{c_{2L-K}} \left(2L-K-g_N(x)\right) \phi(x) dx\nonumber\\
&= \sum_{k=0}^{N} a_k \left[m_k(c_K) + (-1)^k m_k(-c_{2L-K}) \right] - \Phi(-c_K) K - (2L-K)\Phi\left(c_{2L-K}\right) 
\end{align}
with $c_{2L-K}=g_N^{-1}(2L-K)$.

With the latter approach, there is no need for a one-dimensional numerical solver in order to include the martingale constraint into the minimization. The Monte-Carlo simulation of the CLV model becomes however slightly more complex, since there is the need to evaluate the payoff on the mirror path as well.

\subsection{Reflection}
Instead of absorption, we can consider that the asset $S$ reflects at the level $L$, and takes the value 
$2L-g_N(x)$ for $x < g_N^{-1}(L)$. It is still possible to compute the price of a European option analytically. We simply split the integration at $c_L=g_N^{-1}(L)$:
\begin{equation}
C(K) \approx \int_{c_L}^{\infty} \max\left(g_N(x)-K,0\right)\phi(x) dx + \int_{-\infty}^{c_L} \max\left(2L-g_N(x)-K, 0\right) \phi(x) dx\,.
\end{equation}
Assuming that $K > L$, we have
\begin{align}
C(K) &\approx \int_{c_K}^{\infty} \left(g_N(x)-K\right) \phi(x) dx + \int_{-\infty}^{c_{2L-K}} \left(2L-K-g_N(x)\right) \phi(x) dx\nonumber\\
&= \sum_{k=0}^{N} a_k \left[m_k(c_K) - (-1)^k m_k(-c_{2L-K}) \right] - \Phi(-c_K) K + (2L-K)\Phi\left(c_{2L-K}\right) 
\end{align}
with $c_{2L-K}=g_N^{-1}(2L-K)$.
We can still calibrate the model while preserving the first moment. With reflection, the first moment is composed of two parts:
\begin{align}
F(0,T) &= \int_{c_L}^{\infty} g_N(x) \phi(x)dx + \int_{-\infty}^{c_L} \left(2L-g_N(x)\right) \phi(x) dx\,,\\
&=2L \Phi(c_L) + \sum_{i=0}^{N} a_i m_i(c_L) - (-1)^i a_i  m_i(-c_L)\,.
\end{align}

\begin{figure}[h]
	\centering{
		\includegraphics[width=0.95\textwidth]{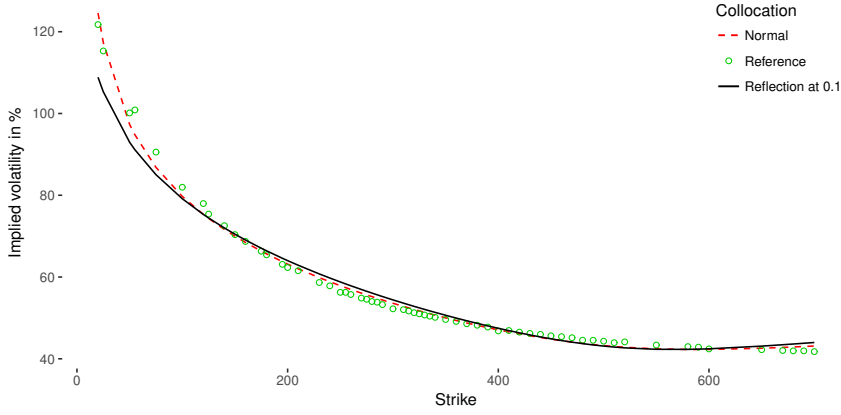}
		\caption{Implied volatility obtained from the calibrated quintic collocation polynomial with reflection for TSLA options expiring on January 17, 2020 as of June 15, 2018.\label{fig:tsla_180615_200117_polyrefl}}}
\end{figure}

 The Monte-Carlo simulation  becomes straightforward as there is no dependency on the reflection at previous times. But the model is absolutely not realistic. And in practice, it does not calibrate very well to the market (Figure \ref{fig:tsla_180615_200117_polyrefl} and Table \ref{tbl:rmse_collocation_tsla}). The preservation of the forward ratio of equation (\ref{eqn:forward_ratio}) also leads to huge autocorrelation corrections with reflection.

\subsection{Exponential extrapolation}
A simpler alternative is to extrapolate the collocation polynomial with an exponential $e^{\alpha x + \beta}$ for $x < x_L$, where $x_L$ is chosen so that $g_N(x_L)$ corresponds to a positive asset price. For example, we can choose the lowest market option strike as extrapolation level. The two parameters $\alpha$ and $\beta$ are chosen so that the collocation is of class $C^1$:
\begin{align}
g_N(x_L) &= e^{\alpha x_L + \beta}\,,\\
g_N'(x_L) &= \alpha e^{\alpha x_L+\beta} = g_N(x_L) \alpha\,,
\end{align}
or equivalently
\begin{align}
\alpha = \frac{g_N'(x_L)}{g_N(x_L)}\,,\\
\beta = \ln g_N(x_L) - \alpha x_L\,.
\end{align}
As $g_N$ is strictly increasing and $g_N(x_L) > 0$ by hypothesis, we have $\alpha > 0$. And thus when $x \to -\infty$, $e^{\alpha x + \beta} \to 0$. The asset can not take a negative value, and all strictly positive values are allowed.

The first moment is now given by
\begin{align}
F(0,T) &= \int_{-\infty}^{x_L} e^{\alpha x + \beta} \phi(x) dx + \int_{x_L}^{\infty} g_N(x) \phi(x) dx\nonumber\\ 
&=e^{\beta+\frac{1}{2}\alpha^2} \int_{-\infty}^{x_L} \phi(x-\alpha) dx + \sum_{i=0}^N a_i m_i(x_L)\nonumber\\
&= e^{\beta+\frac{1}{2}\alpha^2} \Phi(x_L-\alpha) + \sum_{i=0}^N a_i m_i(x_L)\,.
\end{align}
The extrapolation corresponds to a lognormal tail of standard deviation $\alpha$  as 
\begin{equation*}
\int_{-\infty}^{x_L} e^{\alpha x + \beta} \phi(x) dx = \int_{0}^{e^{\alpha x_L + \beta}} z \frac{\phi(\frac{\ln z -\beta}{\alpha})}{\alpha z} dz\,.
\end{equation*} 

With the extrapolation, the undiscounted price of a Call option for $K \geq g_N(x_L)$ is unchanged while the undiscounted price of a Call option for $K < g_N(x_L)$ reads
\begin{equation}
C(K) =  e^{\beta+\frac{1}{2}\alpha^2} \left(\Phi(x_L-\alpha) - \Phi(c_K-\alpha)\right) + \sum_{i=0}^N a_i m_i(x_L) - K \Phi(-c_K)\,,
\end{equation}
where $c_K = \frac{\ln K - \beta}{\alpha}$.
 
The initial guess for the collocation polynomial is the same as without extrapolation. In the non-linear least-squares minimization, the coefficient $a_0$ is implied from the first moment conservation.
\begin{figure}[h]
	\centering{
		\includegraphics[width=0.95\textwidth]{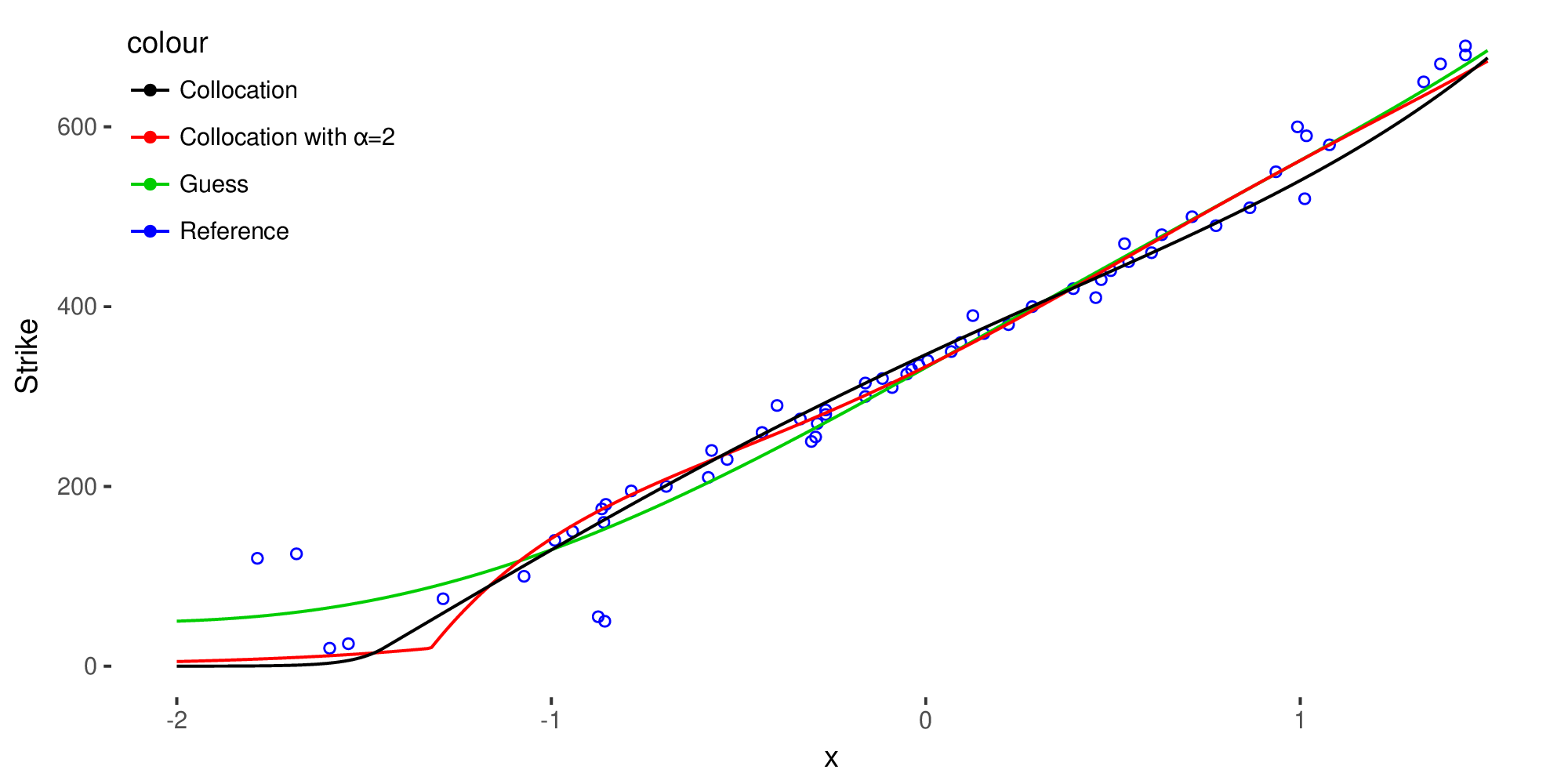}
	\caption{Quintic collocation polynomial with exponential extrapolation for TSLA options expiring on January 17, 2020 as of June 15, 2018.\label{fig:polyextra}}}
\end{figure}

A potential issue is to obtain a too large coefficient $\alpha$. In this case, the autocorrelation implied by the Gaussian CLV model will explode. In practice this can happen when the strike cut-off $g_N(x_L)$ is very low. Two remedies are possible: increase the strike cut-off or cap the coefficient $\alpha$, for example to 2.0. In the latter case, we don't impose the continuity of the first derivative of the collocation function anymore and the density will be discontinuous at the cut-off. Figure \ref{fig:polyextra} shows the quintic collocation polynomial with continuous, or capped extrapolation, calibrated to TSLA options expiring on January 17, 2020.
%
%

\subsection{Lognormal collocation}
Fo most financial asset prices, lognormal returns are a typical modelling assumption. It would thus make more sense, a priori, to collocate towards a log-normal random variable $X$  with standard deviation $\sigma_X$, instead of a normal random variable, using an increasing polynomial map from $[0, \infty)$ to $[0, \infty)$, in a similar fashion as \cite{ware2019polynomial,boonstra2021valuation} do for power prices and the electricity commodity market.

The steps from section \ref{sec:overview_collocation} become: 
\begin{enumerate}
	\item Given a set of collocation strikes $y_i$, $i=0,...,N$, compute the survival density $p_i$ at those points: $p_i = G_Y(y_i)$.
	\item Project on the log-normal distribution by transforming the $p_i$ using the inverse cumulative normal distribution $\Phi^{-1}$ resulting in $x_i = e^{ \sigma_X \Phi^{-1}(1-p_i)}$.
	\item Interpolate $(x_i,y_i)$ with a monotonically increasing polynomial $g_N$.
	\item Price by integrating on the density with the integration variable $x=  e^{\sigma_X\Phi^{-1}(1-G_Y(y))}$, using the Lagrange polynomial for the transform.
\end{enumerate}
Now, the undiscounted price of a vanilla Call option is given by
\begin{align}
C(K) &= \int_{0}^\infty \max\left(G_Y^{-1}\left(1-\Phi\left(\frac{\ln x}{\sigma_X}\right)\right) -K,  0\right) \frac{\phi\left(\frac{\ln x}{\sigma_X} \right)}{\sigma_X x}dx \label{eqn:ln_exact}\nonumber\\
&\approx \int_{0}^{\infty} \max\left(g_N(x) -K , 0\right) \frac{\phi\left(\frac{\ln x}{\sigma_X} \right)}{\sigma_X x}dx \nonumber\\
&= \int_{-\infty}^{\infty} \max\left(g_N\left(e^{\sigma_X y}\right)-K, 0\right) \phi(y) dy\nonumber\\
&= \int_{c_K}^{\infty} \left( g_N\left(e^{\sigma_X y}\right)-K \right) \phi(y) dy\,. 
\end{align}
 where $c_K$ is chosen so that $g_N\left(e^{\sigma_X c_K}\right) = K$. Note that $c_K$ is not guaranteed to exist anymore as $\exists K \in \mathbb{R} | g_N^{-1}(K) \leq 0$. In this case, we take $c_K$ such that $g_N(e^{\sigma_X c_K}) = \epsilon$ with $\epsilon$ a small positive real number, typically the machine epsilon.
 
As in the Gaussian case, the Call price can be computed explicitly:
\begin{align}
C(K) = \sum_{k=0}^N a_k e^{\frac{1}{2}k^2\sigma_X^2} \Phi(k\sigma_X-c_K) - K \Phi(-c_K)\,. 
\end{align}
And the first moment is given by
\begin{align}
F(0,T) &= \int_{-\infty}^{\infty} g_N\left(e^{\sigma_X y}\right) \phi(y) dy\nonumber\\
&= \sum_{k=0}^N a_k \int_{-\infty}^{\infty} e^{k \sigma_X y}\phi(y) dy\nonumber\\
&= \sum_{k=0}^N a_k e^{\frac{1}{2}k^2\sigma_X^2}\,. 
\end{align}

The asset price must stay positive. This implies that $\forall y \in \mathbb{R}, g_N(e^{\sigma_X y}) > 0$. By construction, we choose $g_N$ to be monotonically increasing. And thus we must have $g_N(0)=a_0 \geq 0$. In order to allow for all positive asset prices, we must have $a_0$ = 0.

From the preservation of the first moment, we imply the first degree monomial coefficient $a_1$. The free parameters are then $a_2, a_3, ..., a_N$. Alternatively, we may scale the coefficients $a_i$ by the ratio of market forward price to theoretical forward price. The isotonic representation of \cite{lefloch2019model1} is still applicable, with minor modifications: according to \cite{murray2016fast}, a polynomial $p(x)$ of degree $q$ is non-negative on $[0, \infty)$ if and only if it can be written as 
\begin{equation*}
	p(x) = p_1(x)^2 + x p_2(x)^2\,, \quad \forall x \in \mathbb{R}\,,
	\end{equation*}
where $q=2K$, $p_1(x)$ and $p_2(x)$ are polynomials whose degrees are at most $K$ and $K-1$ respectively, and, if, $q= 2K+1$, both degrees are at most $K$.

In practice however, it turns out not to work very well. A core issue is a lack of invariance in the lognormal collocation.
\begin{definition}
	A polynomial collocation towards a given distribution $F_{X,\sigma}$ is invariant with regards to $\sigma$ if,
given a polynomial $g_1$ and a volatility $\sigma_1$, for any $\sigma_2 > 0$ we can find a collocation polynomial $g_2$ associated to the volatility $\sigma_2$ such that $g_2(x) = g_1(x)$  for all $x$.
\end{definition}
In the case of the normal distribution, $g_2$ is simply a scaling factor multiplied by $g_1$. In the case of the lognormal distribution, the collocation with $\sigma_2 \neq \sigma_1$ is fundamentally different, since the polynomial $g_2$ is applied to $e^{\sigma_2 x}$ while $g_1$ is applied to $e^{\sigma_1 x}$. We can not reconcile the two unless they are of degree zero.
And thus the collocation towards a lognormal variate $X$ will be very dependent on the choice of the volatility $\sigma_X$. In some sense (Equation \ref{eqn:ln_exact}), the basis will effectively be $\left(e^{\sigma_X x}, e^{2 \sigma_X x}, e^{3 \sigma_X x},...\right)$, which does not seem particularly appropriate.

\begin{figure}[h]
	\centering{
		\subfigure[\label{fig:polylog_sigma1} $\sigma_X=1$.]{
			\includegraphics[width=0.45\textwidth]{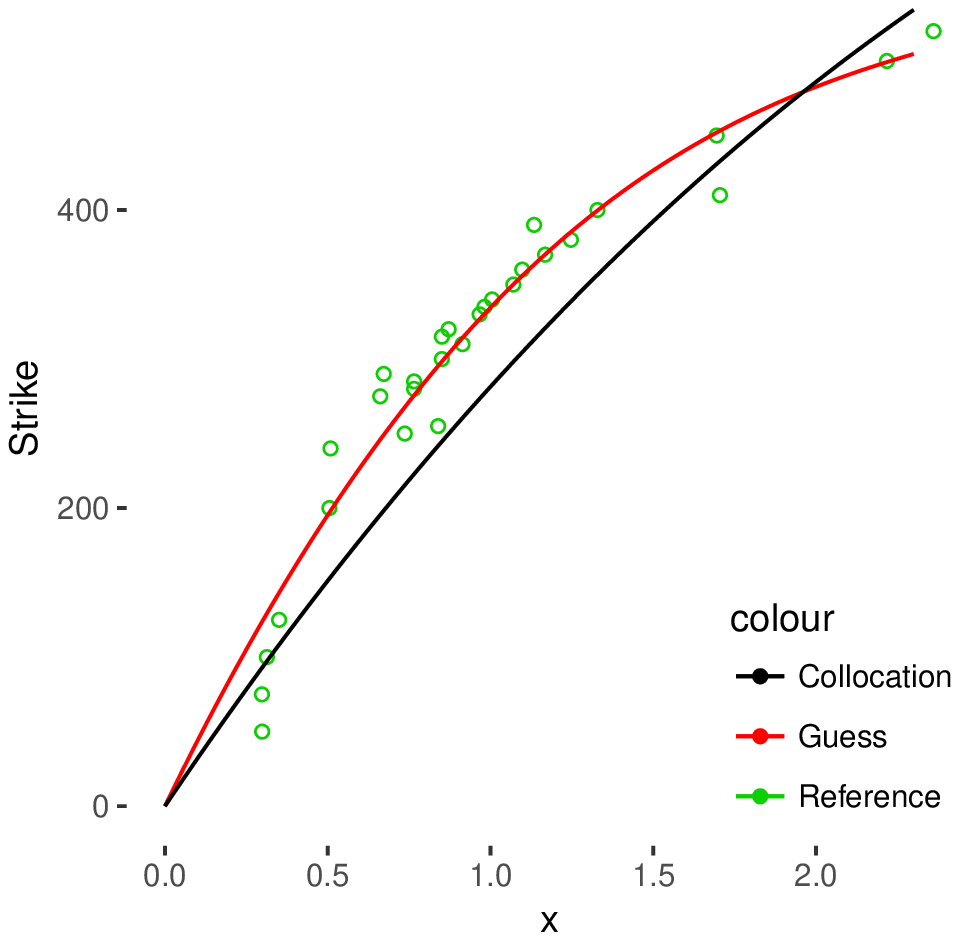}}\hspace{0.05\textwidth}
		\subfigure[\label{fig:polylog_sigma01} $\sigma_X=0.1$.]{
			\includegraphics[width=0.45\textwidth]{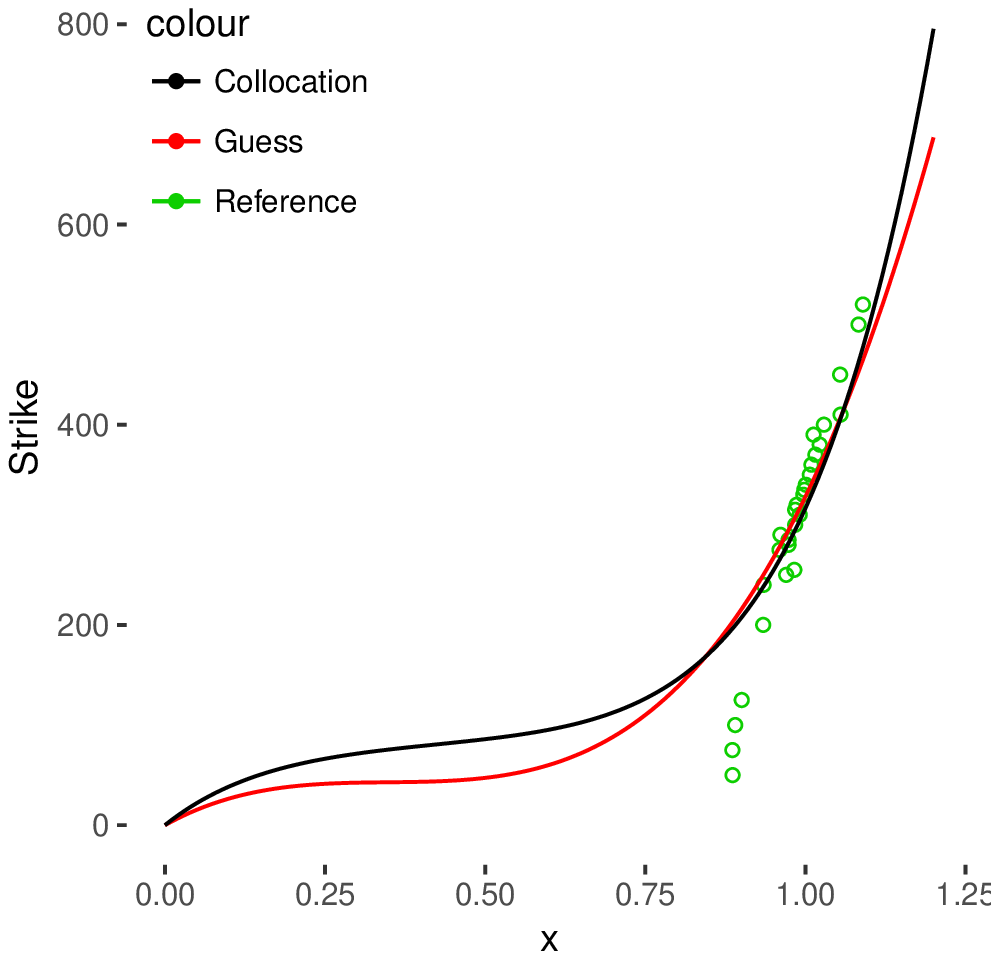} }
	}
	\caption{Quintic collocation polynomial for a lognormal variable on TSLA options expiring on January 17, 2020 as of June 15, 2018.}
\end{figure}

If we take $\sigma_X$ too large, then polynomials will not  approximate properly the density: the left tail of $(x,y)$ behaves like an exponential (Figure \ref{fig:polylog_sigma1}). If we take $\sigma_X$ low, the fit is not better: the slope implied by the market data is very large around $x=1$, which would typically make an unconstrained best fit polynomial go through $(x, 0)$ with $x$ relatively close to 1, but the condition $a_0=0$  forces the polynomial to pass through the origin, a very different constraint (Figure \ref{fig:polylog_sigma01}).
The polynomial collocation towards a lognormal variable is thus not well suited.

\begin{figure}[h]
	\centering{
		\subfigure[\label{fig:polylogn_sigma1} $\sigma_X=1$.]{
			\includegraphics[width=0.45\textwidth]{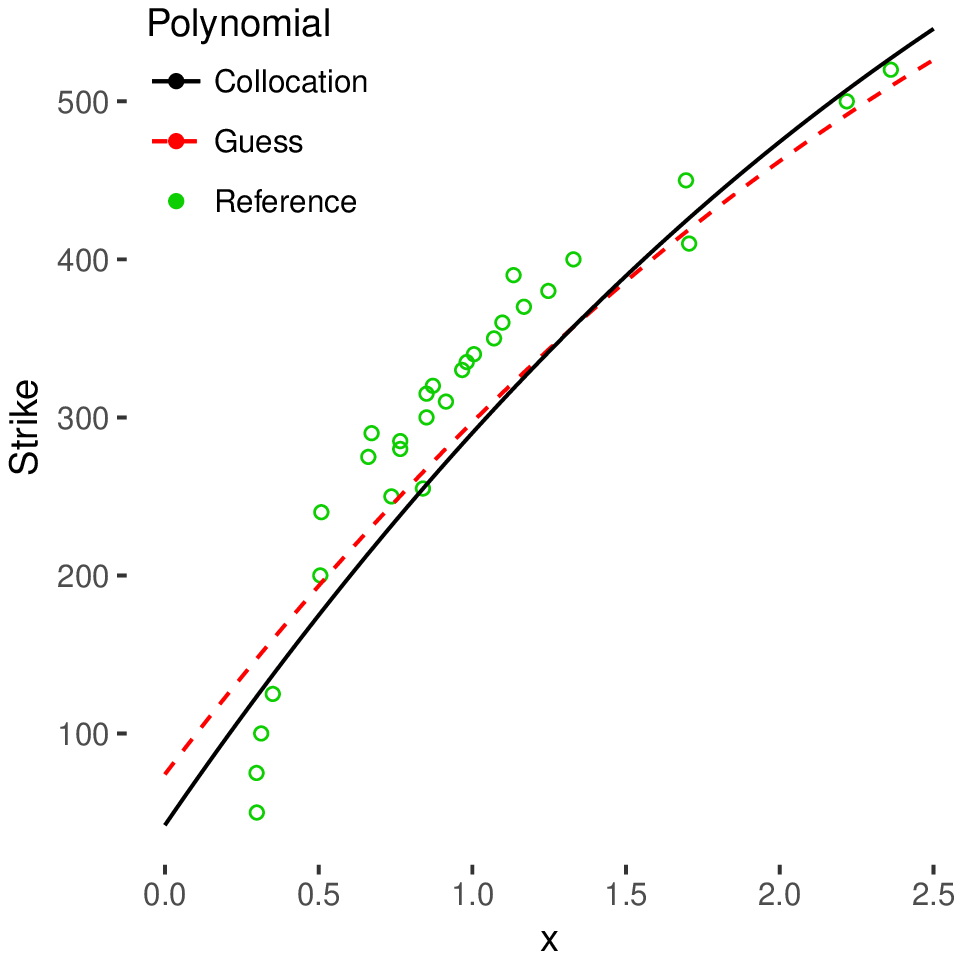}}\hspace{0.05\textwidth}
		\subfigure[\label{fig:polylogn_sigmamin} $\sigma_X=0.547$.]{
			\includegraphics[width=0.45\textwidth]{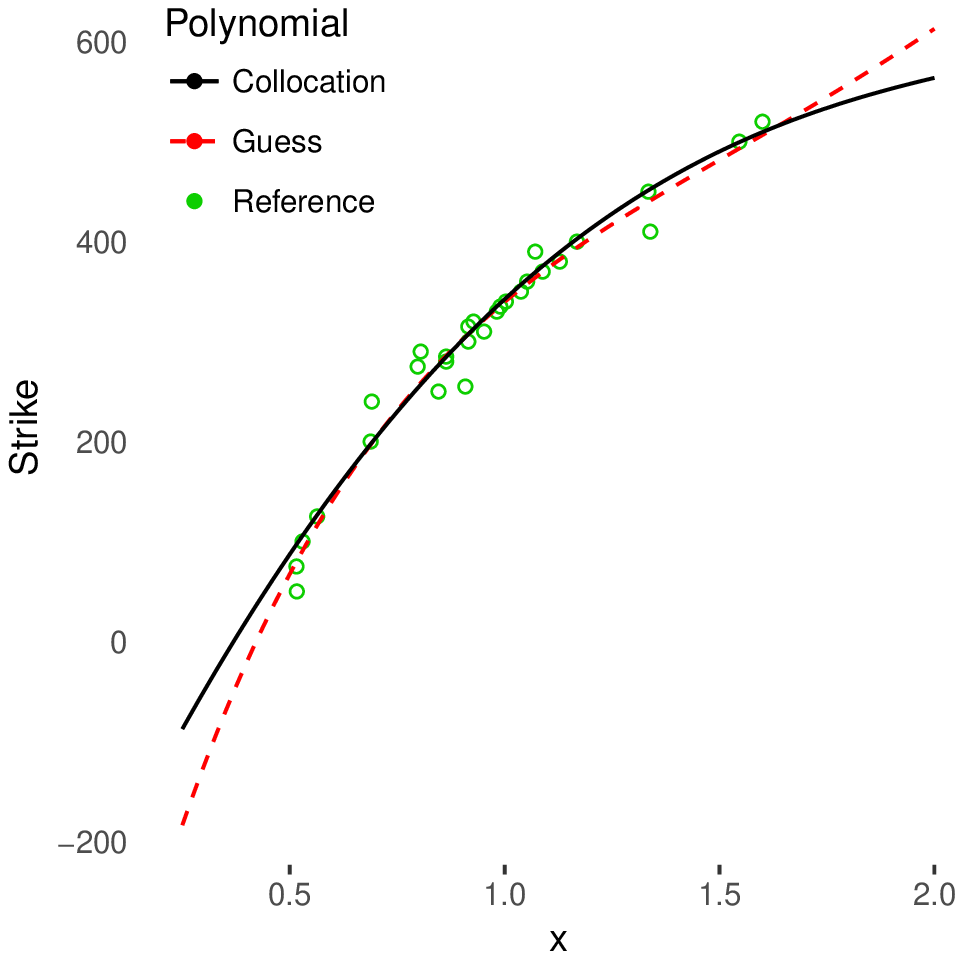} }
	}
	\caption{Quintic collocation polynomial for a lognormal variable on TSLA options expiring on January 17, 2020 as of June 15, 2018, without the constraint $a_0=0$.}
\end{figure}

\begin{figure}[h]
	\centering{
	\includegraphics[width=0.95\textwidth]{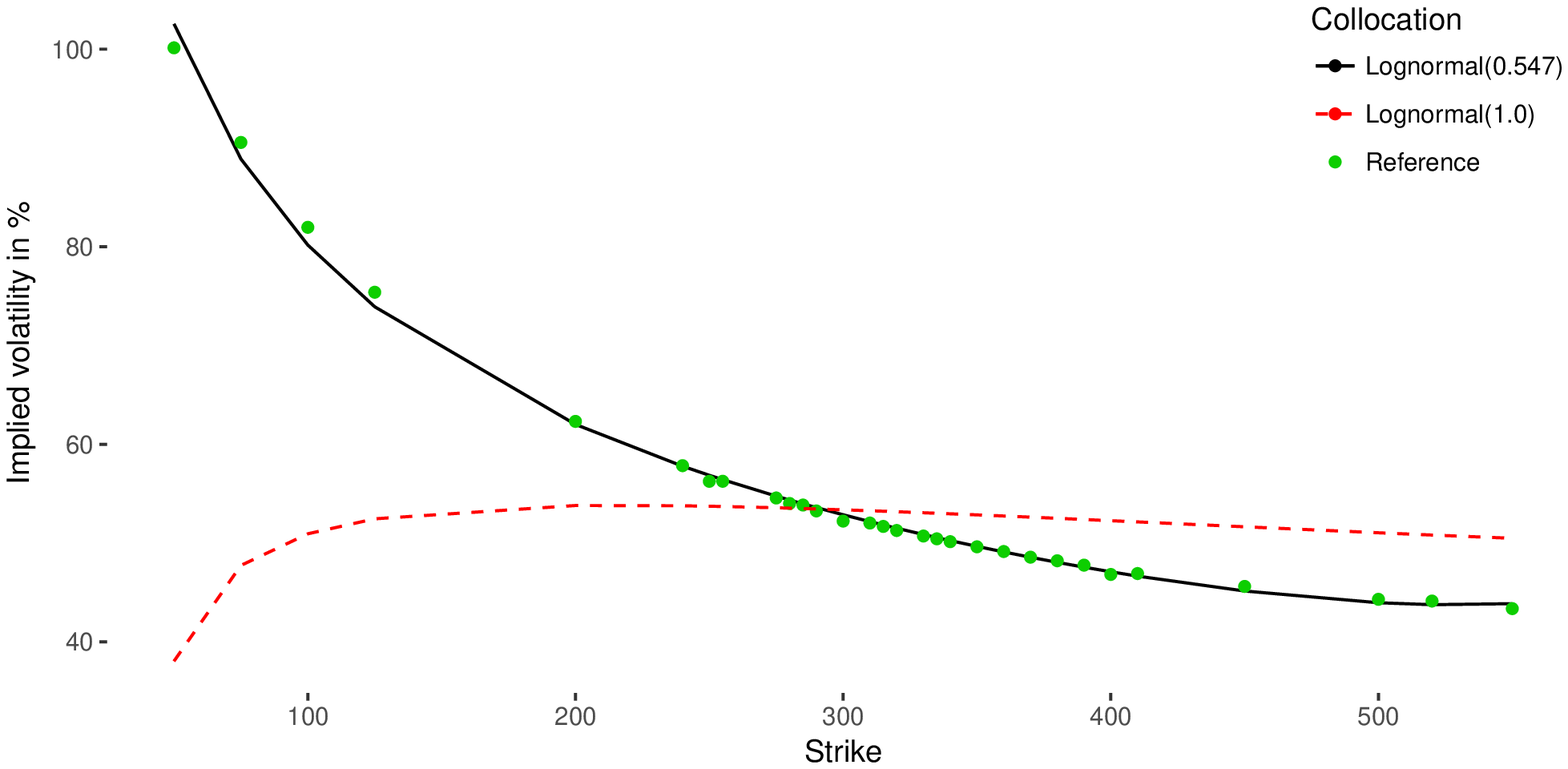}
	\caption{Implied volatility obtained from the calibrated unconstrained quintic lognormal collocation polynomial for TSLA options expiring on January 17, 2020 as of June 15, 2018, for two different values of $\sigma_X$.\label{fig:tsla_180615_200117_polylogn}}}
\end{figure}

If we relax the positivity constraint and allow $a_0 \neq 0$, then the lognormal collocation can fit the market data much better, as long as $\sigma_X$ is not too large. In figure \ref{fig:polylogn_sigma1} and \ref{fig:polylogn_sigmamin}, we consider $\sigma_X=1$ and $\sigma_X = \min_{i=0,...,m}(\sigma_i \sqrt{T}) \approx 0.547$ where $\sigma_i$ is the market implied volatility for the option of strike $K_i$.
When $\sigma_X$ is too large ($\sigma_X=1$ in our example), the unrestricted polynomial still can not fit the market data points. When $\sigma_X=0.547$, a polynomial can fit reasonably well the market data. In terms of implied volatilities, the fit is acceptable (Figure \ref{fig:tsla_180615_200117_polylogn}).

In order to ensure the positivity of the asset price under collocation, without the constraint $a_0=0$, we can add an exponential extrapolation, as in the standard normal collocation. The first moment becomes:
\begin{equation}\label{eqn:logextra_collo_martingale}
F(0,T) = e^{\beta + \frac{1}{2}\alpha^2} \Phi(c_L-\alpha) + \sum_{k=0}^N a_k e^{\frac{1}{2}k^2\sigma_X^2} \Phi(k\sigma_X-c_L)\,,
\end{equation}
where $L$ is a positive cut-off strike and $c_L = \frac{1}{\sigma_X}\ln g_N^{-1}(L)$.
And when $K < L$, the call option price $C$ becomes
\begin{equation}
C(K) = e^{\beta + \frac{1}{2}\alpha^2}\left( \Phi(c_L-\alpha) -  \Phi(c_K-\alpha) \right)+  \sum_{k=0}^N a_k e^{\frac{1}{2}k^2\sigma_X^2} \Phi(k\sigma_X-c_L) - K \Phi(-c_K)\,, 
\end{equation}
with  $c_K = \frac{ \ln(K)-\beta}{\alpha}$.

The extrapolation parameters $\alpha$ and $\beta$ are determined so that the extrapolation is of class $C^1$:
\begin{align}
g_N\left(e^{\sigma_X c_L}\right) &= e^{\alpha c_L + \beta}\,,\\
\sigma_X e^{\sigma_X c_L} g_N'\left(e^{\sigma_X c_L}\right) &= \alpha e^{\alpha c_L + \beta}\,,
\end{align}
or equivalently
\begin{align}
\alpha &= \sigma_X e^{\sigma_X c_L}\frac{g_N'\left(e^{\sigma_X c_L}\right)}{g_N\left(e^{\sigma_X c_L}\right)}\,,\\
\beta &= \ln g_N\left(e^{\sigma_X c_L}\right)  - \alpha c_L\,.
\end{align}
As $g_N$ is strictly monotone and increasing by hypothesis, we have $g_N' > 0$. Furthermore, we choose $c_L$ so that the corresponding strike $L=g_N\left(e^{\sigma_X c_L}\right)$ is strictly positive, and the standard deviation $\sigma_X$ is strictly positive, thus we have $\alpha > 0$.

In order to impose the martingality condition, we solve for the coefficient $a_0$ in equation (\ref{eqn:logextra_collo_martingale}), with a one-dimensional root solver such as Toms348 \citep{alefeld1995algorithm}.

\begin{figure}[h]
	\centering{
		\includegraphics[width=0.95\textwidth]{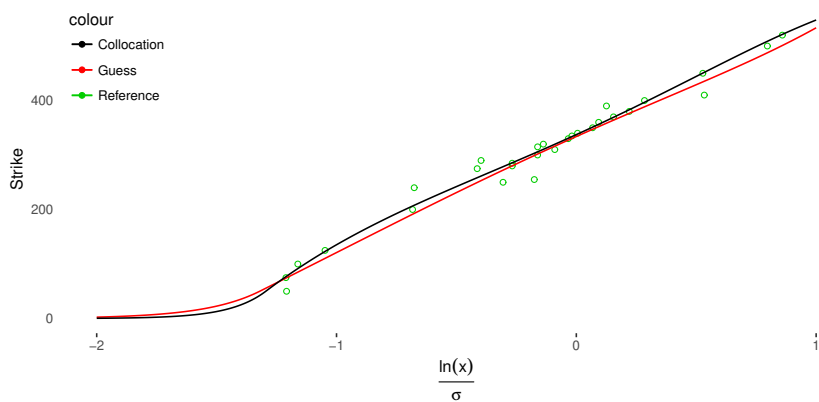}
		\caption{Quintic lognormal collocation polynomial with exponential extrapolation for TSLA options expiring on January 17, 2020 as of June 15, 2018. The final lognormal standard deviation is chosen in order to minimize the least-squares error in option prices, with an initial lognormal standard deviation $\sigma=0.547$.\label{fig:polylognextra}}}
\end{figure}

Instead of fixing the standard deviation $\sigma_X$, we can use it as an additional calibration parameter in the least squares minimization. This can significantly improve the fit (Figure \ref{fig:polylognextra}). For the TSLA options expiring on January 17, 2020, the optimal value is $\sigma_X \approx 0.223$. 

\begin{table}[h]
	\caption{Root mean square error (RMSE) in implied volatilities obtained by various collocation methods on TSLA options of maturity January 17, 2020. \label{tbl:rmse_collocation_tsla}}
	\centering{
		\begin{tabular}{lr}\toprule
			Collocation method & RMSE\\ \midrule
 Normal     & 0.0142\\
 Absorption & 0.0073\\
 Reflection  &0.0302\\
 Exponential extrapolation with optimal $\alpha=13.4$ & 0.0064\\
 Exponential extrapolation with capped $\alpha=2$ & 0.0077\\
 Lognormal with min. vol $\sigma=0.527$ & 0.0227\\
 Lognormal with optimal $\sigma=0.144$ & 0.0173 \\
  Lognormal with min. vol $\sigma=0.527$ and optimal $\alpha=11.3$ & 0.0067\\
 Lognormal with optimal $\sigma=0.091$ and optimal $\alpha=13$ & 0.0052\\
 Lognormal with optimal $\sigma=0.0545$ and capped $\alpha=2$ & 0.0177\\
			\bottomrule
	\end{tabular}}
\end{table}

\section{Calibration of the autocorrelation in the CLV model}\label{sec:calibration_autocorrelation_clv}
\subsection{The drift in the CLV model}
With an arithmetic Brownian driver process \footnote{This extends to more sophisticated processes such as the  Ornstein-Uhlenbeck process}, the CLV model from \citet{grzelak2016clv} does not respect
\begin{equation}\label{eqn:forward_ratio}
\mathbb{E}_{\mathbb{Q}}\left[\frac{S_i}{S_j}\right] = \frac{F_i}{F_j}\,.
\end{equation}
It is thus not compatible with an assumption of  deterministic interest rates. In fact, when the asset is used as the numeraire, the value $\mathbb{E}_{\mathbb{Q}}\left[\frac{S_i}{S_j}\right]$ implies an inherent stochastic interest rate behaviour, for which there is reason to be in line with the market \citep{fries2006markov}. 

The presence of a spurious drift can be deduced from the application of the It\^o Lemma:
\begin{align}
dS = \left(\frac{\partial g}{\partial t}+ \frac{\sigma_X(t)^2}{2}\frac{\partial^2 g}{\partial x^2}\right)dt + \sigma_X(t)\frac{\partial g}{\partial x}dW(t)\,,
\end{align}
the drift is not zero in general. \citet{carr1999closed} have shown that imposing $\frac{\partial g}{\partial t}+ \frac{\sigma_X(t)^2}{2}\frac{\partial^2 g}{\partial x^2}=0, \forall t>0, \forall x \in \mathbb{R}$, is a very restrictive condition on $g$.
\subsection{Calibration of the autocorrelation}
Similarly to the Markov functional model, for a given discretization, it is possible to force a specific autocorrelation of the driver process in order to match $\mathbb{E}\left[\frac{S_i}{S_j}\right] = \frac{F_i}{F_j}$, as long as the process $S$ is guaranteed to stay strictly positive.

With the driver $X(t)=\sigma_X(t)W(t)$, for $t_i < t_j$, assuming that $\sigma_X(t)>0$, the autocorrelation reads
\begin{equation}
\rho_{i,j} = \frac{\mathbb{E}\left[X(t_i)X(t_j)\right]}{\sqrt{\int_{0}^{t_i} \sigma_X(u)^2 du \int_{0}^{t_j} \sigma_X(u)^2 du}} = \sqrt{\frac{\int_{0}^{t_i} \sigma_X(u)^2 du}{\int_{0}^{t_j} \sigma_X(u)^2 du}}\,.\label{eqn:autocorrelation_sigmat}
\end{equation}
In particular, for  a Brownian motion, when $\sigma_X$ is constant,  we have \begin{equation*}\rho_{i,j}=\sqrt{\frac{t_i}{t_j}}\,.\end{equation*}
Instead of this autocorrelation, as in \citep{jackel2005practical,brockhaus2006implied}, we choose a specific autocorrelation $\rho_{i,j}$ such that the deterministic forward ratio is preserved:
\begin{align}
\mathbb{E}_{\mathbb{Q}}\left[\frac{S_j}{S_i}\right] &= \frac{F(0,t_j)}{F(0,t_i)}\,.
\end{align}
The expectation can be computed in closed form. We have
\begin{align}
\mathbb{E}_{\mathbb{Q}}\left[\frac{S_j}{S_i}\right] &= \int_{-\infty}^{\infty} \int_{-\infty}^{\infty} \frac{g(t_j,\rho_{i,j} u+\sqrt{1-\rho_{i,j}^2}v)}{g(t_i,v)} \phi(u) \phi(v) du dv\,.\label{eqn:autocorrelation_collocation}
\end{align}
If we integrate first on the variable $u$, the inner integral has an explicit formula in terms of the normal cumulative function $\Phi$ in a similar fashion as for the first moment.
For the Gaussian collocation with exponential extrapolation, we obtain:
\begin{align}
\mathbb{E}_{\mathbb{Q}}\left[\frac{S_j}{S_i}\right] &= \int_{-\infty}^{\infty} \frac{e^{\frac{1}{2}\alpha^2\bar{\rho}_{i,j}^2+ \alpha \rho v +\beta}\Phi\left(c(v)-\alpha\bar{\rho}_{i,j}\right)}{g(t_i,v)}\phi(v)  dv \nonumber \\
&+\int_{-\infty}^{\infty}  \frac{\sum_{k=0}^N a_k(t_j) \sum_{l=0}^k {k\choose l} \bar{\rho}_{i,j}^{l} \rho_{i,j}^{k-l} v^{k-l}m_l(c(v)) }{g(t_i,v)} \phi(v)  dv\,,\label{eqn:autocorrelation_collocation_polyextra}
\end{align}
with $c(v) = \frac{c_L-\rho_{i,j} v}{\bar{\rho}_{i,j}}$, $\bar{\rho}_{i,j} = \sqrt{1-\rho_{i,j}^2}$, $c_L= g^{-1}(t_j,L)$, $L$ is the extrapolation cut-off strike, and $a_k(t_j)$ are the coefficient of the $k$-th monomial of $g_N(t_j)$.
The outer integral can be computed with an adaptive Gauss-Lobatto quadrature.

We can thus calibrate the $\rho_{i,j}$ for $t_i < t_j$ to deterministic forward ratios. The constant Gaussian driver autocorrelation can be used as initial guess.

	 Alternatively, from equation (\ref{eqn:autocorrelation_sigmat}), we could also calibrate a piecewise constant volatility parameter $\sigma_X(t)$ approximately to the deterministic forward ratios (in this case there are $m$ constants to calibrate against $m\cdot\frac{m-1}{2}$ forward ratios, where $m$ is the number of time-steps).


The autocorrelations lead to the following correlation matrix $A$ relating normal variates to the path values on the normalized coordinate $x$:
\begin{equation}
A \cdot \begin{pmatrix}
z_1\\
z_2 \\
z_3\\
\vdots\\
z_m
\end{pmatrix}
=\begin{pmatrix}
x_1 \\
x_2 \\
x_3\\
\vdots\\
x_m
\end{pmatrix}\,.
\end{equation}
with \begin{equation}
A A^T = \begin{pmatrix}
1 &   \rho_{1,2}   & \rho_{1,3} & \dots&  \rho_{1,m}\\
\rho_{1,2} & 1 & \rho_{2,3} & \dots&  \rho_{2,m}\\
\rho_{1,3} & \rho_{2,3} & 1 & \dots & \rho_{3,m} \\
\vdots &                &    &   \ddots   & \vdots\\          
\rho_{1,m} & \rho_{2,m} & \rho_{3,m} & \dots & 1
\end{pmatrix}
\end{equation}
The correlation matrix exists only if the covariance matrix is positive definite and can be computed by Cholesky decomposition or by singular value decomposition (SVD). The latter allows to compute an reasonably good approximative correlation matrix even if the original covariance matrix is not positive definite \citep{jackel2004monte}.

In the Monte-Carlo simulation, we then compute each path according to the following steps:
\begin{itemize}
	\item Draw $m$ independent uniform random numbers $U_i$.
	\item	Transform them to Gaussian random numbers $Z_i = \Phi^{-1}(U_i)$.
	\item	Compute $X_i = \sum_{j=1}^{m} A_{i,j} Z_j$. 
	\item Compute $S(t_i) = g_N(t_i, X_i)$.
\end{itemize}

\subsection{A refined criteria}\label{sec:refined_criteria}
\citet{brockhaus2006implied} proposes a more general identity than the forward ratio in the absence of arbitrage. In order to match the behavior with deterministic interest rates, we must have
\begin{equation}
\mathbb{E}_{\mathbb{Q}}\left[\frac{S_j}{S_i} 1_{S_i > B}\right]  = \frac{F(0,t_j)}{F(0,t_j)} P(S_i > B)\,,
\end{equation}
where $B$ is a barrier level, and $P(S_i > B)$ is the probability that the asset $S$ is above $B$ at time $t_i$. 
With the CLV model, this can still be computed in closed form:
\begin{align}
\mathbb{E}_{\mathbb{Q}}\left[\frac{S_j}{S_i} 1_{S_i > B}\right] &= \int_{c_B}^{\infty} \int_{-\infty}^{\infty} \frac{g(t_j,\rho_{i,j} u+\sqrt{1-\rho_{i,j}^2}v)}{g(t_i,v)} \phi(u) \phi(v) du dv\,,\label{eqn:autocorrelationb_collocation}
\end{align}
where $c_B = g^{-1}(t_i,B)$. And, for the standard collocation, we also have
\begin{equation}
P(S_i > B) = \Phi(-c_B)\,,
\end{equation}
for $B \geq L$ in the case of an extrapolation cut-off at $L$.

\subsection{Numerical example on TSLA options}
\begin{figure}[h]
	\centering{
		\includegraphics[width=0.95\textwidth]{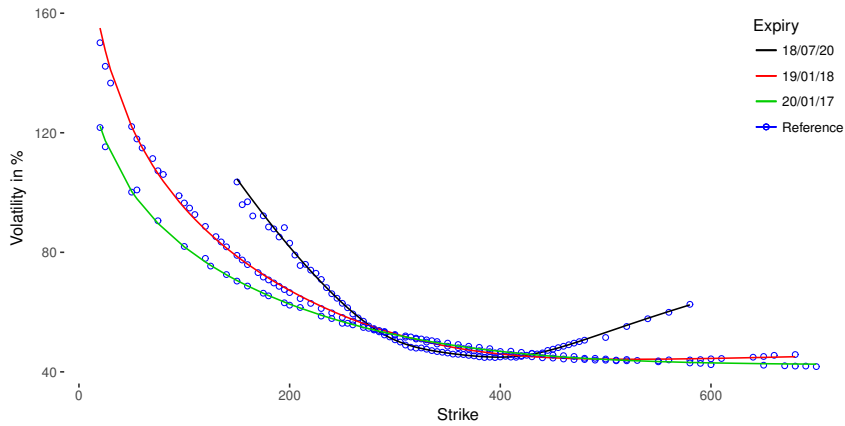}
		\caption{Implied volatility smiles obtained by quintic collocation with exponential extrapolation for TSLA options as of June 15, 2018.\label{fig:tsla_180615_polyexra_vols}}}
\end{figure}

On the 15th of June 2018, we consider options on the TSLA stock at three distinct maturities: July 20, 2018 (1 month), January 18, 2019 (7 months), January 17, 2020 (19 months).
We calibrate quintic collocation polynomials with exponential extrapolation to this data. This results in the implied volatility smiles of Figure \ref{fig:tsla_180615_polyexra_vols}. The fit for each option maturity is excellent.

In table \ref{tbl:autocorrelations_polyextra}, we then compute the optimal autocorrelations to preserve the deterministic forward ratios between each maturity according to equation (\ref{eqn:autocorrelation_collocation}).
\begin{table}[h]
	\caption{Autocorrelations between option maturities. The indices 1, 2, 3 denote the first, second and third option maturities. \label{tbl:autocorrelations_polyextra}}
	\centering{
		\begin{tabular}{lrrr}\toprule
			Autocorrelation method & $\rho_{1,2}$ & $\rho_{1,3}$ & $\rho_{2,3}$ \\ \midrule
			Wiener & 0.4016 & 0.2454 & 0.6111 \\
			Calibrated & 0.4017 & 0.2531 & 0.9061 \\
			Calibrated with $\alpha$ capped to 2& 0.4021 & 0.2505 & 0.7743 \\			
			\bottomrule
	\end{tabular}}
\end{table}

Finally, in table \ref{tbl:mc_forwardratio_polyextra}, we verify with a Monte-Carlo simulation that the calibrated autocorrelations allow to preserve the forward ratios, and compare with the raw autocorrelations of a Wiener process with constant volatility.
\begin{table}[h]
	\caption{Monte-Carlo simulation of the contract on the ratio $\frac{S(t_3)}{S(t_2)}$ with 4 million paths. The deterministic value is 1.0017 \label{tbl:mc_forwardratio_polyextra}}
	\centering{
		\begin{tabular}{lrrr}\toprule
			Autocorrelation method & Monte-Carlo mean (std. err) &  Error & Analytical value \\ \midrule
& \multicolumn{3}{c}{$\alpha$ uncapped}\\
			Wiener & 13484.071 ($\pm$ 4911)&  13483.0700 & $2.74\cdot {10}^7$\\ 
			Calibrated & 0.966 ($\pm$ 0.007) &  -0.0360 & 1.0017\\\cmidrule(lr){2-4}
& \multicolumn{3}{c}{$\alpha$ capped to 2}\\
			Wiener & 1.1606 ($\pm$ 0.0013)&  0.1589 & 1.1607\\
Calibrated & 1.0017 ($\pm$ 0.0005) &  0.0000 & 1.0017\\
			\bottomrule
	\end{tabular}}
\end{table}
In particular, we observe a significant drift of the forward without
calibration of the autocorrelations. This is much more pronounced without cap on the $\alpha$ extrapolation parameter in the collocation polynomial calibration, but is still important with a relatively low cap. The autocorrelations adjustment does not work as well for the uncapped exponential extrapolation, because with a large extrapolation factor $\alpha$, the extrapolation is very close to zero for relatively small negative values of $x$. This creates an instability in the Monte-Carlo simulation as the asset almost reaches zero too quickly.

The CLV model with adjusted autocorrelation does not lead to an easy PDE representation. It is much simpler to consider a term-structure of volatilities for the driver process instead, which will allow to match approximately the forward ratios between each expiries.

We calibrate a term structure of piecewise constant forward volatilities for the driver process, instead of the direct individual autocorrelation, with only three maturities. Let the forward volatility be $\sigma_{k}$ between $t_{k}$ and $t_{k+1}$, the total variance up to a time $t_i$ is then
\begin{equation}
\int_0^{t_i} \sigma_X^2(u) du = \sum_{k=1}^{i} \sigma_{k-1}^2 (t_k-t_{k-1})\,,
\end{equation} 
with $t_0 = 0$. In particular, it is guaranteed to increase, and thus the model is well defined, in contrast to the approach of \cite{brockhaus2006implied}. The choice of the first volatility parameter $\sigma_0$ is free, we take simply $\sigma_0=1.0$ and calibrate $\sigma_1$ and $\sigma_2$ to the forward ratios $\mathbb{E}_{\mathbb{Q}}\left[\frac{S_2}{S_1}\right]$, $\mathbb{E}_{\mathbb{Q}}\left[\frac{S_3}{S_1}\right]$, $\mathbb{E}_{\mathbb{Q}}\left[\frac{S_3}{S_2}\right]$.

With the capped extrapolation, we find a solution that is close to the three forward ratios (Table \ref{tbl:forwardvol_polyextra}).
\begin{table}[h]
	\caption{Forward volatilities calibrated to the forward ratios. $e_{i,j}$ is the error in the forward ratio $\mathbb{E}_{\mathbb{Q}}\left[\frac{S_j}{S_i}\right]$ \label{tbl:forwardvol_polyextra}.}
	\centering{
		\begin{tabular}{rrrrrr}\toprule
			$\sigma_1$ &  $\sigma_2$ & $e_{1,2}$ & $e_{1,3}$ & $e_{2,3}$ \\ \midrule
			1.154 & 0.714 & 0.0030 & -0.0025 & 0.0001\\
			\bottomrule
	\end{tabular}}
\end{table}
On our example, there are two forward volatilities to calibrate to three forward ratios. In the more general case, there are $n-1$ forward volatilities to calibrate to $\frac{n(n-1)}{2}$ forward ratios, and thus with more expiries, the error in the forward ratios is likely to be larger.

\begin{figure}[h]
	\centering{
		\includegraphics[width=0.95\textwidth]{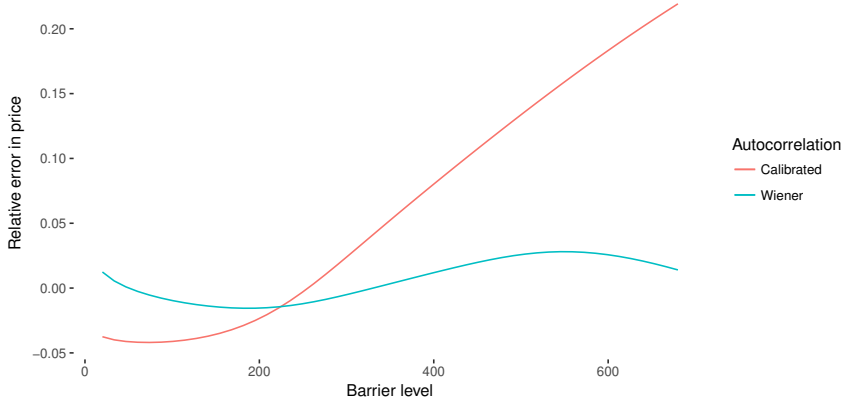}
		\caption{Relative error in the forward performance $\frac{S_3}{S_2}$ with barrier at $B$, varying $B$ for the collocation with capped extrapolation.\label{fig:tsla_polyextra_ratiob}}}
\end{figure}

With the calibrated autocorrelations, we can also evaluate the error of the refined criteria described in section \ref{sec:refined_criteria}.
On Figure \ref{fig:tsla_polyextra_ratiob}, we plot the relative error in the price of the forward performance $\frac{S_3}{S_2}$ with barrier at $B$, varying $B$.
For relatively low strikes, the error can be of around 5\%  and for large strikes, it can be even higher. This is in line with \citet{brockhaus2006implied} observations, and is the drawback of the CLV or the Markov functional models applied to equities. 
When the autocorrelation is not calibrated, the relative error in the  forward performance with barrier is actually smaller, below 3\%. But without barrier, the relative error reaches $16\%$ as shown in Table \ref{tbl:mc_forwardratio_polyextra}. 
Regarding the effect of autocorrelation and the calibration to the forward performance, the lognormal collocation behaves the same way as the collocation towards a Gaussian random variable.

In comparison, the local volatility model is fully consistent, but offers no control over forward smile dynamic, which is unrealistic.

\section{Local volatility for a term-structure of collocated smiles}\label{sec:local_volatility_collocation}
\citet{gatheral2014arbitrage} present the discrete version of the calendar spread no-arbitrage condition:
\begin{equation}
\frac{C\left(K_2,T_2\right)}{K_2 }\ \geq \frac{C\left(K_1,T_1\right)}{K_1} 
\end{equation}
for $T_2 \geq T_1$, where $C$ is the undiscounted call option price
and the strikes $K_2, K_1$ are chosen so that the forward moneyness is constant: 
\begin{equation*}
\frac{K_1}{F(0,T_1)} = \frac{K_2}{F(0,T_2)}\,.
\end{equation*}
This can be derived from Jensen inequality, knowing that the process $\frac{S(t)}{F(0,t)}$ is a martingale.

This discrete arbitrage-free inequality can be used to define an arbitrage-free interpolation in time assuming that it is verified at two expiries $t_1$ and $t_2$. For example, we can define the prices at time $t_1 < t < t_2$ by
\begin{align}
C\left(K,t\right) &= \frac{t-t_1}{t_2-t_1} \frac{F(0,t)}{F(0,t_2)} C\left(\frac{K F(0,t_2)}{F(0,t)},t_2\right)+  \frac{t_2-t}{t_2-t_1} \frac{F(0,t)}{F(0,t_1)} C\left(\frac{K F(0,t_1)}{F(0,t)},t_1\right)\,.
\end{align} 

\citet{dupire1994pricing} expresses the local volatility $\sigma_L$ as:
\begin{equation}
\sigma_L^2(K,T) = 2 \frac{\frac{\partial C_0}{\partial T}}{K^2\frac{\partial^2 C_0}{\partial K^2}}\,,
\end{equation} 
where $C_0(K,T)$ is the undiscounted call option price on a fixed forward $F(0,T)$. With the collocation method, the Dupire local volatility has a closed form expression with the above interpolation of Call prices in time.
\begin{equation}
\frac{\partial C_0}{\partial T}(K,t) = \frac{K}{t_2-t_1} \left( \frac{C(K_2,t_2)}{K_2}-\frac{C(K_1,t_1)}{K_1}\right)
\end{equation}
for $\frac{K}{F(0,t)}= \frac{K_1}{F(0,T_1)} = \frac{K_2}{F(0,T_2)}$.

\begin{equation}
\frac{\partial^2 C_0}{\partial K^2}(K,t) = \frac{t-t_1}{t_2-t_1} \frac{K_2}{K} \frac{\partial^2 C}{\partial K^2}\left(K_2,t_2\right)+  \frac{t_2-t}{t_2-t_1} \frac{K_1}{K}  \frac{\partial^2 C}{\partial K^2}\left(K_1,t_1\right)\,.
\end{equation}

When the call option prices are determined by the collocation method with a function $g_N$, we have
\begin{equation}
\frac{\partial C}{\partial K}(K_i)= - \Phi(-c_{K_i})\,,\quad 
\frac{\partial^2 C}{\partial K^2}(K_i) = \frac{\phi(c_{K_i})}{{g_i}_N'(c_{K_i})} \,,
\end{equation}
with $c_{K_i} = {g_i}_N^{-1}\left(K_i\right)$.
For $t_1 < t < t_2$, the Dupire equation becomes
\begin{equation}
\sigma_L(K,t) = 2K^2\frac{ \frac{C(K_2,t_2)}{K_2}-\frac{C(K_1,t_1)}{K_1}  }{ (t-t_1)K_2 \frac{\phi(c_{K_2})}{{g_2}_N'(c_{K_2})} + (t_2-t)K_1 \frac{\phi(c_{K_1})}{{g_1}_N'(c_{K_1})} }\,.
\end{equation}

\section{Conclusion}
Among the various ways to apply the stochastic collocation technique to represent the prices of vanilla options on an asset constrained above boundary, we found that absorption and exponential extrapolation were preferable, for their simplicity and their ability to fit the market, while the polynomial collocation towards a lognormal distribution was much less adequate in practice, partly because of its strong dependency on the choice of lognormal volatility.

We also explained that the CLV model does not respect some key identities linked to the future performance under the assumption of deterministic interest rates, as well as the amplitude of the discrepancies. While this could be considered as a barrier to the use of such a model, it is not necessarily worse than the popular Dupire local volatility model. The Dupire local volatility model is consistent with regards to those identities, but suffers from a bad forward smile dynamic, and will thus give particularly bad prices for contracts written on a future performance. The CLV model allows some flexibility towards the capture of this forward smile dynamic, which could be more in line with the market of forward starting (or barrier) options, but suffers from those other discrepancies.

Finally, we have shown that the stochastic collocation technique can also be interesting to use as a basis to obtain a smooth Dupire local volatility, for example in the context of a stochastic volatility model.
\bibliographystyle{spbasic}      

\bibliography{lefloch_collocation}

\end{document}